\documentclass[journal = nalefd, manuscript = letter, layout = traditional]{achemso}

\usepackage{amssymb}
\usepackage{amsmath}
\usepackage{graphicx}
\usepackage{changes}
\usepackage{color}

\author{Nojoon Myoung}
\affiliation{Department of Physics Education, Chosun University, Gwangju 61452, Republic of Korea}
\author{Hyungkook Choi}
\affiliation{Department of Physics, Chonbuk National University, Jeonju 54896, Republic of Korea}
\author{Hee Chul Park}
\email{hcpark@ibs.re.kr}
\affiliation{Center for Theoretical Physics of Complex Systems, Institute for Basic Science, Daejeon 34126, Republic of Korea}

\title{Manipulation of valley isospins in strained graphene for valleytronics}


\begin{document}

\begin{tocentry}
\centering
\includegraphics[height=3.5cm]{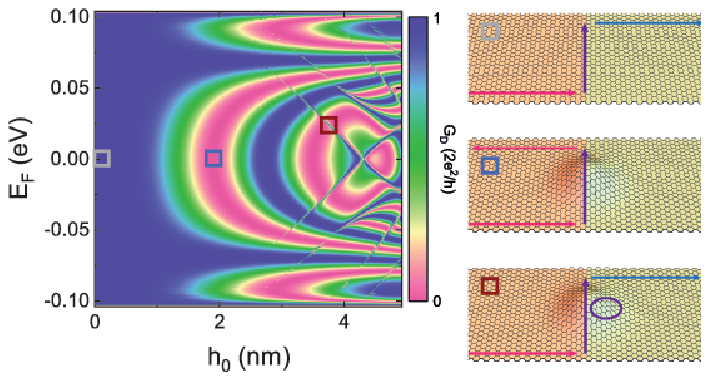}
\end{tocentry}

\begin{abstract}
Graphene's outstanding mechanical properties lend to strain engineering, allowing for future valleytronics and nanoelectromechanic applications. In this work, we have found that a Gaussian-shaped strain on a graphene p--n junction results in quantum Hall conductance oscillations due to the rotated angle between valley isospins at the graphene armchair edges. Furthermore, additional Fano resonances were observed as the value of the strain-induced pseudo-magnetic field approaches that of the external magnetic field. The lifted valley degeneracy, stemming from the interplay between the real and pseudo-magnetic fields, results in clearly valley-resolved Fano resonances. Exploring strain engineering as a means to control conductance through valley isospin manipulation is believed to open the door to potential graphene valleytronic devices.
\end{abstract}

Graphene is a fascinating material in both fundamental research and applications for its unique electrical, mechanical, and optical properties.\cite{Novoselov2005,Geim2007,Balandin2008,Lee2008,Bolotin2008,Kim2009,Li2009}
With carrier mobilities reaching up to a few million cm$^{2}$V$^{-1}$s$^{-1}$,\cite{Banszerus2015} pristine graphene has been subjected to a myriad of studies as a potential candidate to extend beyond traditional silicon-based electronic technologies; much work has been focused on applications such as spintronics,\cite{Hill2006,Tombros2007,Haugen2008,Dlubak2012,Han2014,Berger2015} optoelectronics,\cite{Bonaccorso2010,Gu2012,Britnell2013,Yu2013,Zhang2014,Kuiri2016,Zhang2017} plasmonics,\cite{Ju2011,Koppens2011,Grigorenko2012,Low2014,Rodrigo2015} and sensors.\cite{Cho2015,Rodrigo2015,Choi2017} Likewise, the so-called valleytronics\cite{GarciaPomar2008,Gunlycke2011,Wu2011,Jiang2013} aspects of graphene in particular have recently been attracting attention---manipulaton of the valley degree of freedom as a key knob to control electric current, like as spin in spintronics. The most promising path to graphene valleytronics has been reported to be strain engineering\cite{Fujita2010,Low2010,Wu2011,Si2016} due to the extraordinary mechanical properties of graphene.

In the tight-binding approach, strain in graphene is regarded as a synthetic gauge field and a pseudo-magnetic field, which are of opposite signs for Dirac fermions near different valleys.\cite{Low2010,Guinea2010,Vozmediano2010,deJuan2012} Interplay between the pseudo-magnetic field and an external magnetic field has been suggested as a way to manipulate the valley degree of freedom.\cite{Kim2011} Various realizations of strain control in graphene have been reported,\cite{Xu2012,Lu2012,Klimov2012,Jiang2013a,Smith2016} with such strain tunability hinting at the feasibility of practical graphene-based valleytronic devices.

In general, valleys can be mixed by the presence of boundary conditions, as at graphene edges. For instance, valleys are fully mixed for armchair graphene nanoribbons, while they are polarized for zigzag graphene nanoribbons.\cite{Brey2006,Park2011} Such valley-mixing behavior is well characterized by introducing the concept of \textit{valley isospin} in a Bloch sphere as if describing the singlet spin state.\cite{Tworzdlo2007,Recher2007,Liang2008,Beenakker2008} In the coherent regime, valley-isospin dependence has been revealed to be significant, influencing quantum Hall conductance across a p--n junction in graphene nanoribbons.\cite{Tworzdlo2007} Recently, a cutting-edge experiment has confirmed valley-isospin dependence by identifying the widths of graphene nanoribbons with atomic-scale precision.\cite{Handschin2017} Interest in the valley isospin of graphene nanoribbons keeps increasing with such experimental progress.

In this Letter, we present theoretical calculations that show how the valley-isospin dependence of quantum Hall effects in graphene is influenced by elastic strain. Conductance is measured through a one-dimensional interface channel formed at a p--n junction, and a Gaussian-shaped local strain is applied to the vicinity of the junction. We first explore the strain effects on conductance by varying the strength and position of the local strain, resulting in conductance oscillation as a consequence of valley-isospin rotation. We demonstrate that valley isospin is rotated by the phase Dirac fermions acquire while traveling along the interface channel. In addition, we investigate the properties of localized states in the strained region when the strength of the strain-induced pseudo-magnetic field becomes comparable to the external magnetic field. We reveal that the existence of these localized states leads to Fano resonances in the conductance, and furthermore, that the localized states are valley-resolved due to the interplay between the pseudo- and external magnetic fields.

\section{Model}

\begin{figure}
\includegraphics[width=8.5cm]{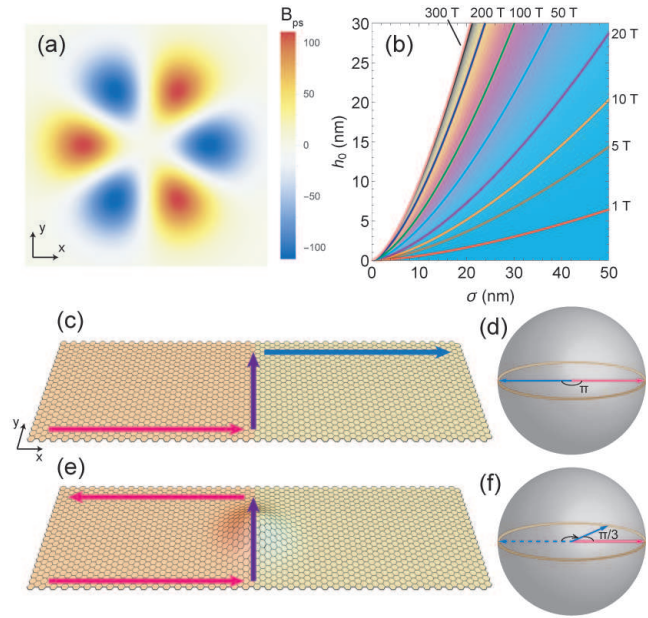}
\caption{(a) Pseudo-magnetic field profile induced by a Gaussian deformation of the K valley. (b) Maximum strength of the pseudo-magnetic field as a function of the size and height of deformation, denoted by $\sigma$ and $h_{0}$, respectively. Solid curves imply constant pseudo-magnetic field lines. (c) and (e) Schematic views of the system without and with strain. (d) and (f) Valley-isospin configurations in the absence and presence of strain. Blue and red arrows indicate the valley isospins at each edge of the system. } \label{fg:model}
\end{figure}

The graphene quantum Hall bar in this work is subjected to a Gaussian-shaped deformation that results in local strain, $z\left(\vec{r}\right)=h_{0}\mbox{exp}{\left[-\left(\vec{r}-\vec{r}_{0}\right)^{2}/\left(2\sigma^{2}\right)\right]},$ where $h_{0}$ is the maximal deformation in the vertical direction $z$ at the center, i.e., at $\vec{r}=\vec{r}_{0}$, and $\sigma$ is the standard deviation. In the case of unrelaxed lattices, Gaussian deformation possesses out-of-plane strain only.\cite{Moldovan2013} Deformation results in differing inter-carbon distances in the strained region, which in turn causes changes in the hopping energies between adjacent sites, from the tight-binding point of view. As a result, it has been found that the pseudo-magnetic field is given by\cite{Schneider2015}
\begin{align}
\vec{\mathcal{B}}_{ps}=\nu\frac{\hbar\beta}{ea_{0}}\frac{h_{0}^{2}}{\sigma^{6}}e^{-\frac{r^{2}}{\sigma^{2}}}r^{3}\sin{3\theta}\hat{z},
\end{align}
where $\beta=3.37$ and $a_{0}=0.142$ nm is the inter-carbon distance of graphene. The pseudo-magnetic field satisfies $\vec{\mathcal{B}}_{ps}=\vec{\nabla}\times\vec{\mathcal{A}}$ where $\vec{\mathcal{A}}$ is the strain-induced gauge field (see the Supplementary Material for details of the physical model). A profile of the pseudo-magnetic field with Gaussian deformation is displayed in Fig. \ref{fg:model}(a), where it can be noticed that the field acts oppositely on different valleys. To compare its strength with the external magnetic field, it is convenient to find the maximum magnitude of the pseudo-magnetic field, $\mathcal{B}_{ps,max}=\nu e^{-3/2}\left(27\hbar\beta h_{0}^{2}\right)/\left(8ea_{0}\sigma^{3}\right),$ which is plotted in Fig. \ref{fg:model}(b) as a function of $\sigma$ and $h_{0}$.

Now, we discuss the electrostatic potential distribution in a p--n junction, where interface channels are created. In this Letter, we consider a gently varying potential profile to allow for practical fabrications in experimental studies. Note that coherent transport through the lowest Landau level (LL) channel is not influenced by the quality of the junction profile\cite{LaGasse2016,Myoung2017} (see Supplementaty Material for details of the junction profile). For simplicity, the p--n junction is anti-symmetrically produced and only the lowest-LL channels are taken into account.

For graphene nanoribbons, it has been shown that quantum Hall conductance across a p--n junction depends on the orientation angle between valley isospins at the edges in each region.\cite{Tworzdlo2007} In the presence of strain, the valley-isospin dependence of quantum Hall effects in armchair graphene nanoribbons can be well formulated by
\begin{align}
G_{D}=\frac{G_{0}}{2}\left[1-\cos{\left(\Phi+\Phi_{ps}\right)}\right],
\end{align}
where $G_{0}=2e^{2}/h$, $\Phi$ is the angle between valley isospins, and $\Phi_{ps}$ is the net phase acquired from the gauge fields along the interface channel. $G_{D}$ is the quantum Hall conductance across the p--n junction, measured via diagonal leads as depicted in Fig. \ref{fg:model}(c) and (e). It is worth mentioning that $\Phi_{ps}$ is continuously given, whereas $\Phi$ is given by three-fold values: $\pi$ or $\pm\pi/3$ for metallic and semiconducting cases, respectively. In this Letter, we set $\Phi=\pi$. When a Gaussian-deformation is created near the p--n junction, the valley isospin rotates as a consequence of phase acquirement $\Phi_{ps}$, such that $G$ varies. The goal of this work is to propose a feasible way of manipulating valley isospins through the strain engineering of graphene.

Our theoretical methodology to calculate the transport properties of the given system is as follows. The quantum Hall conductance is calculated from S-matrix formalism with the tight-binding approach using the KWANT code,\cite{Groth2014} and a hopping energy of $3.0$ eV is used in the tight-binding calculation.

\section{Results and Discussion}

\begin{figure*}
\includegraphics[width=16cm]{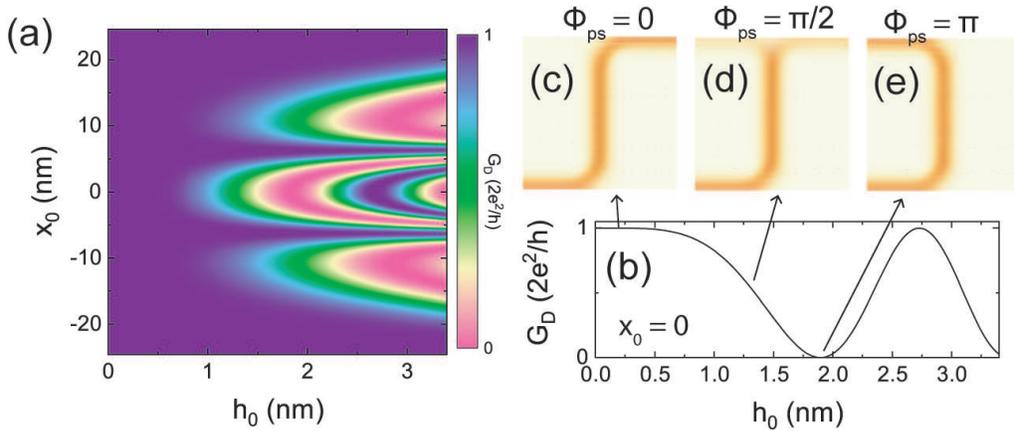}
\caption{ (a) Dependence of conductance across p--n junction $G_{D}$ on the location $x_{0}$ and maximum height $h_{0}$ of the Gaussian deformation for $\sigma=$ 7.4 nm. (b) $G_{D}$ versus $h_{0}$ for $x_{0}=0$. (c), (d), and (e) Probability current density maps of three selected values of conductance: $G_{D}=G_{0}$, $0.5~G_{0}$, and 0, respectively, with corresponding values of $h_{0}$ given as 0, 1.3, and 1.8 nm, respectively. } \label{fg:strain}
\end{figure*}

Figure \ref{fg:strain} shows how the conductance across junction $G_{D}$ is influenced by Gaussian deformation. One can clearly see that $G_{D}$ remains unchanged for sufficiently small $h_{0}$ since the strain-induced phase acquirement is not large enough. Meanwhile, the conductance exhibits an oscillatory behavior as the position of the deformation varies. Since Gaussian deformation creates nonuniform pseudo-magnetic fields (see Fig. \ref{fg:model}(a)), the total phase acquirement of Dirac fermions can differ according to where the p--n interface is formed. Obviously, $G_{D}$ is barely influenced by strain created sufficiently far from the p--n junction.

The strain-induced conductance oscillation versus $h_{0}$ supports the prediction that the angle between the valley isospins rotates when a local strain is given near the p--n junction. The net phase acquirement of Dirac fermions is calculated by integrating the strain-induced pseudo-magnetic fields over the area enclosed by snake-like trajectories along the junction.\cite{Myoung2017} Additionally, one can note that $G_{D}$ is independent of $h_{0}$ for $x_{0}\simeq \pm6$ nm. Such insensitivity to strain is attributed to the fact that the net phase acquirement of Dirac fermions approaches zero (see Supplementary Material for details).

\begin{figure*}
\includegraphics[width=16cm]{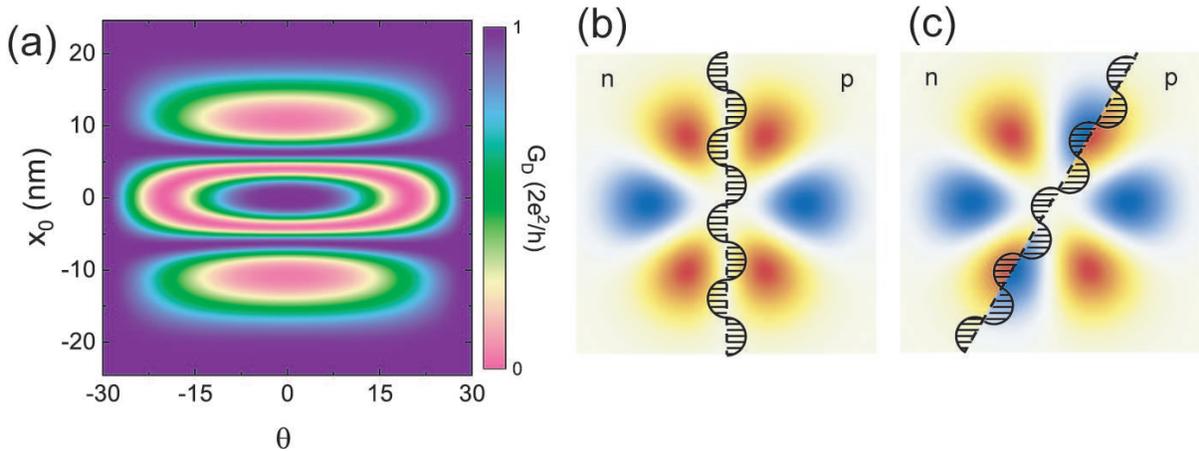}
\caption{ (a) Orientation-angle dependence of conductance across p--n junction $G_{D}$ as $x_{0}$ varies for $\sigma=$ 7.4 nm and $h_{0}=1.6$ nm. (b) and (c) Schematic diagrams for $\theta=0$ and $30^{\circ}$ cases, respectively. The shaded regions enclosed by the snake-like trajectories indicate areas of phase acquirement.} \label{fg:strain2}
\end{figure*}

Since the pseudo-magnetic fields induced by the Gaussian deformation are angular dependent (but periodic with respect to 60$^{\circ}$),$G_{D}$ also exhibits angular dependence with 60$^{\circ}$ periodicity. Indeed, Fig. \ref{fg:strain2}(a) shows that $G_{D}$ has an isotropic angular dependence and exhibits oscillatory behavior as $\theta$ varies. Even though the deformation remains unchanged, net phase acquirement varies as $\theta$ changes, resulting from the nonuniform $B_{ps}$ profiles (see Supplemental Material for details). As mentioned, $G_{D}$ is found to be insensitive to strain for $x_{0}\simeq 6$ nm as a consequence of the vanishing net phase acquirement. Let us now point out an interesting case---when $\theta=30^{\circ}$, $G_{D}$ is completely unchanged by strain, as if the Dirac fermions do not experience the deformation. The existence of such a ``blind'' angle can be understood by examining the extent of phase acquirement while Dirac fermions travel through the p--n interface. By comparing Fig. \ref{fg:strain2}(b) and (c), one can see that the net phase acquirement for $\theta=30^{\circ}$ vanishes because of the antisymmetric $B_{ps}$ profile with respect to the p--n interface, contrary to other cases.

\begin{figure*}
\includegraphics[width=16.0cm]{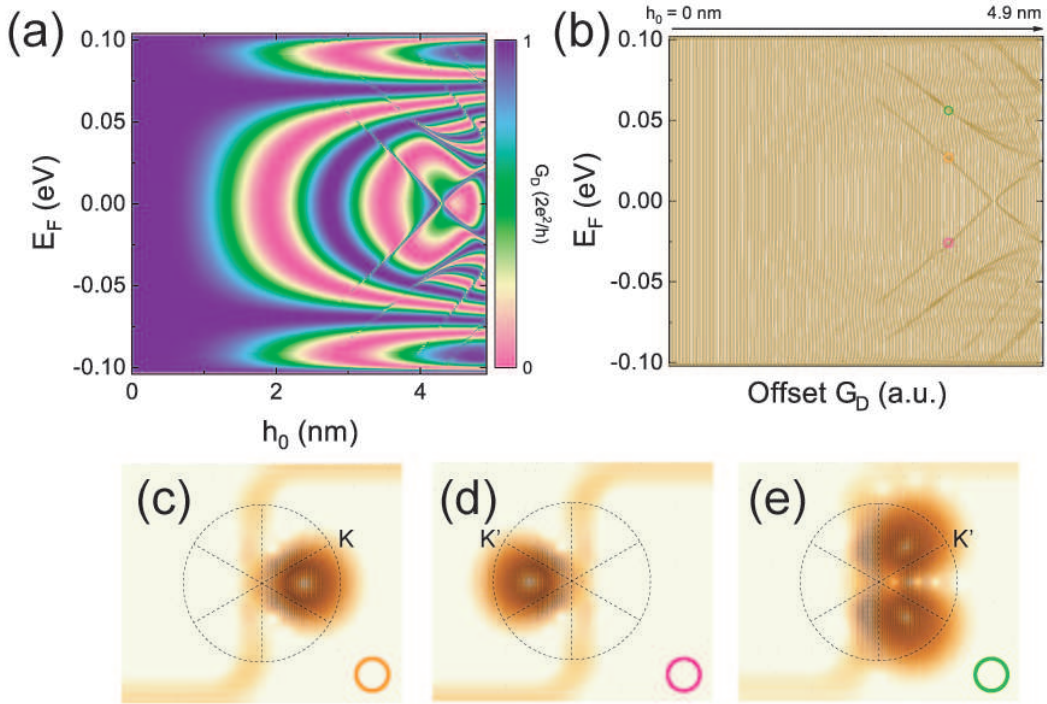}
\caption{ (a) Color map of $G_{D}$ for $\sigma=7.4$ nm as a function of $E_{F}$ and $h_{0}$. (b) Conductance spectra versus $E_{F}$ for various $h_{0}$ from 0 to 4.9 nm with 0.025-nm steps. (c)--(e) Probability current density maps for different Fano resonance lines denoted by orange, red, and green circles in (b). Dashed lines are eye-guides indicating the strained regions divided into six areas with alternating pseudo-magnetic field distributions. } \label{fg:fano}
\end{figure*}

Next, one can see that conductance resonances appear for relatively stronger strain cases, as demonstrated by the diagonal lines along the $G_{D}$ map in Fig. \ref{fg:fano}(a). Owing to the anti-symmetry of the p--n junction, the $G_{D}$ and Fano resonance lines exhibit isotropic spectra with respect to $E_{F}$. These resonances can be understood as consequences of quantum interference between the extended states in the interface channel and the localized states in the strained region,or so-called Fano resonance. The diagonal resonance lines imply that the energy levels of the localized states in the strained region are dependent on $h_{0}$: with increasing strain strength, Dirac fermions more strongly localize, so that the energy levels of the localized states shift from the lowest LLs in each region ($E_{F}\simeq\pm0.1$ eV). 

Meanwhile, as shown in Fig. \ref{fg:fano}(b), there are two distinct types of resonance lines denoted by orange/pink and green circles. Probability density maps for each case are given in Fig. \ref{fg:fano}(c)--(e), showing single- and double-site localizations. These localized states are regarded as if a single quantum dot or double quantum dots are formed in the strained region. Such emergence of strain-induced quantum dots can be understood by seeing how the effective potential is shaped (see Supplemental Material for details). It is also noticeable that a crossing behavior is observed when two single-dot resonances intersect at $E_{F}=0$ eV for $h_{0}=4.2$ nm, whereas an anti-crossing behavior is observed when the single- and double-dot resonances meet near $E_{F}=0.025$ eV for $h_{0}=4.9$ nm. Such crossing/anti-crossing behaviors originate from the valley-polarization of the localized states in strained graphene.\cite{Kim2011} The single-dot localized states shown in Fig. \ref{fg:fano}(c) and (d) correspond to the K and K' valleys, respectively, while the double-dot localized state in Fig. \ref{fg:fano}(e) corresponds to the K' valley. With a lack of inter-valley scattering in this study, it is straightforward to see the crossing behavior between the single-dot resonances because of their opposite valley polarizations. For the same reason, it is clear that the anti-crossing behavior  between the single- and double-dot resonances comes from their identical valley polarizations.

Lastly, let us emphasize the experimental feasibility of our physical model. With $\sigma=$ 21 nm, we notice that $h_{0}$ is limited to about 4.2 nm, not exceeding the practical limit for elastic deformation (20 $\%$). Although the conductance spectra beyond $h_{0}=4.2$ nm in Fig. \ref{fg:fano} are not actually observable, theoretical examination of this regime is helpful for understanding the properties of the strain-induced localized states in this study. Also, as aforementioned, we continuously vary electrostatic potential through the 15 nm distance between the n and p regions for more realistic applications. We stress that the observation of Fano resonances in the conductance spectra is secured even if the slope of the potential profile changes (see Supplementary Material), and finally, we roughly estimate the spectral widths of the Fano resonances to few-meV, which are experimentally distinguishable magnitudes with current techniques.\cite{Weisz2012}

\section{Summary and Conclusions}

In summary, we have studied the influence of local strain on quantum Hall conductance across a p--n interface in graphene. We revealed that the valley-isospin dependence of the quantum Hall conductance is modulated by the presence of local strain near the interface channel. Results indicated that quantum Hall conductance across the p--n interface no longer exhibits a clear plateau but rather an oscillating behavior with respect to strain strength. Such conductance oscillations originate from the rotations of the valley isospins because of the strain-induced pseudo-magnetic field. We have theoretically demonstrated that this valley-isospin rotation indeed occurs by the phase acquired by Dirac fermions while traveling through the strained region. Finally, we discussed the emergence of Fano resonances as evidence for the existence of localized states in a local strain. The strain-induced localized states are regarded as valley-resolved quantum dots in either single or double form.

Our findings in this work give rise to two significant implications in the field of graphene-based valleytronics. First, the conductance oscillation due to strain-induced valley-isospin rotation delivers a realizable approach to manipulate valley isospins in graphene quantum Hall devices, which means that the transport properties of such devices could be controllable via strain engineering. Second, the emergence of Fano resonances with the valley-resolved localized states possesses a great deal of potential for a novel type of valleytronic application based on graphene. Furthermore, the self-assembled localized states in the strained region may open an efficient means of fabricating a perfectly symmetric configuration of double or triple quantum dots.

\section*{Author Information}

\subsection*{Corresponding Author}

E-mail: hcpark@ibs.re.kr

\subsection*{Author Contribution}

The manuscript was written through contributions of all authors. All authors have given approval to the final version of the manuscript.

\section*{Note}

Any additional relevant notes should be placed here.

\begin{acknowledgement}
This work is supported by the National Research Foundation of Korea (NRF) grant and Korea Institute for Advanced Study(KIAS) funded by the Korea government (MSIT) (2017R1C1B5076824, 2019R1F1A1051215,  Project IBS-R024-D1, 2019R1A5A1012495, 2017R1C1B3004301).
\end{acknowledgement}

\bibliography{StrainQHGra}

\end{document}